\documentclass[conference]{IEEEtran}
\usepackage{calc}

\pdfoutput = 1
\usepackage{graphicx}
\usepackage{caption}
\usepackage{subcaption}

\begin{document}

\title{Mobility Study for Named Data Networking in Wireless Access Networks}

\author{\IEEEauthorblockN{ Aytac Azgin, Ravishankar Ravindran, Guoqiang Wang}
\IEEEauthorblockA{Huawei Research Center, Santa Clara, CA, USA}
\{aytac.azgin,ravi.ravindran,gq.wang\}{@huawei.com}}

\maketitle

\begin{abstract}
Information centric networking (ICN) proposes to redesign the Internet by replacing its host-centric design with information-centric design \cite{ICNforest}. Communication among entities is established at the naming level, with the receiver side (referred to as the \emph{Consumer}) acting as the driving force behind content delivery, by interacting with the network through Interest message transmissions. One of the proposed advantages for ICN is its support for mobility, by de-coupling applications from transport semantics. However, so far, little research has been conducted to understand the interaction between ICN and mobility of consuming and producing applications, in protocols purely based on information-centric principles, particularly in the case of NDN. In this paper, we present our findings on the mobility-based performance of Named Data Networking (NDN) in wireless access networks. Through simulations, we show that the current NDN architecture is not efficient in handling mobility and architectural enhancements needs to be done to fully support mobility of \emph{Consumer}s and \emph{Producer}s.
\end{abstract}

\section{Introduction}

Current Internet architecture relies on the host-centric networking principle and requires end-to-end sessions to be associated with unique host identifiers or addresses. As a result of shifting user interests and online activity models (\emph{i.e.}, putting emphasis on the information rather than the way information is acquired), current Internet technologies have been exposed in their inefficiency of handling the continually increasing user demands (which requires service scaling and migration), end point dynamism (\emph{i.e.}, mobility), and security requirements. To present a better solution to these emerging architectural problems, new networking models have been proposed that emphasize on information-centric content delivery and networking (for a detailed overview see \cite{SurveyICN}).

In \emph{Information-centric Networking} (ICN) the key component and the main driving force for information dissemination is the named-content\footnote{Hereafter, we use content and named-content interchangeably.}. Using a unique naming convention for the content, users make a request for each piece of content to the network, which are then forwarded towards the content custodians (\emph{e.g.}, routers, storage servers, end points) hence not limited to the origin servers. To exploit these \emph{in-network cache}s, ICN relies heavily on multicast mechanisms that support efficient and timely delivery of content to requesting consumers.

However, despite these potential advantages, ICN is still in the research stages with limited implementations to date. Certain aspects of ICN, such as mobility in wireless access networks, have not been deeply investigated yet. For instance, in the case of mobility, even though related concerns have been addressed numerous times in recent proposals and research papers, the level of impact mobility can have on a realistic implementation of the ICN architecture has not been studied yet, neither analytically nor with simulations \cite{SurveyMobility}. In this paper, we aim to fill this gap by presenting a realistic performance study for mobility in Named Data Networks (NDN) using a recently developed simulation framework, ndnSIM \cite{ndnSIM}.

In short, in this paper, our objective is to study the potential problems introduced by user mobility in information-centric networks, with specific emphasis on NDN \cite{ndn,Jacobson}. For this purpose, we develop a simulation framework based on ndnSIM to analyze the impact of mobility on user perceived performance. We focus on the delivery of delay-sensitive and delay-tolerant traffic and consider multiple network topologies to represent host mobility within a single \emph{Autonomous System} (AS) and multiple ASs. Our results show that, using unmodified NDN architecture, end-users experience significant performance degradations due to mobility, and the performance degradation is often unacceptable to satisfy the minimum service quality requirements. We also show that performance limitations introduced due to host mobility severely undermines the advantages presented by features such as in-network caching, and introduce significant overhead to the system based on the content routing strategy.

The rest of the paper is organized as follows. In Section~\ref{Section:NDN} we briefly explain the networking operation in NDN. Section~\ref{Section:Mobility} presents the mobility framework used in our analysis. We present our results in Section~\ref{Section:Performance}. We discuss the implications of mobility in ICN in Section~\ref{Section:Discussions}. Section~\ref{Section:Conclusions} concludes our paper.

\section{Named Data Networking Architecture}\label{Section:NDN}

Architectures for information-centric networks are uniquely defined by how they handle \emph{naming} and \emph{name resolution} \cite{ICNsurvey12}. In NDN, names are structured hierarchically (for scalable routing), which can contain any number of components of arbitrary length. Name resolution is conducted in an online manner with request/response type message (referred to as \emph{Interest} and \emph{Data}) processing at every hop, along the path between the \emph{Consumer} to the points of the content source.

NDN allows end or intermediate nodes to locally broadcast, multicast, or unicast Interests. For instance, in a wireless access context, consumers can perform local broadcast over all available network interfaces (or \emph{faces} in short\footnote{In NDN, \emph{face} is a general term that represents the interface over which the data is received or delivered (\emph{i.e.}, network interface or application interface). Each network face is associated with a set of metrics that allow the host to make the proper forwarding decision based on the implemented policy.}), which is then forwarded (or locally multicast/broadcast depending on name routing information availability) hop-by-hop towards the content source. Any node that has access to the requested content (\emph{e.g.}, router, end user, or origin server) can respond to the received Interest by forwarding the Data packet along the reverse path. By limiting the propagation of Interests towards the origin server, less resources are required to acquire the content, thereby leading to significant bandwidth cost and energy savings in the network. Furthermore, network stability is ensured by allowing each Interest to generate a single Data packet in response.

To support content discovery and delivery in NDN, each node is equipped with three core components:
\begin{itemize}
    \item \emph{Content Store} (CS) represents the local cache for the stored content in an NDN router. Anytime a host receives an Interest message for a content cached in its CS, the host creates the Data response and forwards it through the incoming interface(s) before discarding the Interest.
    \item \emph{Pending Interest Table} (PIT) corresponds to the set of active (or pending) Interest messages forwarded by the host and waiting for the corresponding Data packets to be delivered. Each PIT entry tracks the incoming faces for the Interest messages and is represented at minimum with the tuple ``\emph{\{name, in-face(s)\}}". PIT entries are created per content rather than per request, \emph{i.e.}, any subsequent Interest for an active PIT entry is suppressed at the local host, and the current PIT entry is updated with the incoming face information. Since the PIT size is limited, each entry in the PIT is associated with a timeout that is based on the round-trip-time (RTT) estimate. PIT entry timeouts give a local host the ability for quick path recovery by retransmitting the Interest messages over other available faces. Anytime a host receives a matching Data packet for a pending Interest, received Data packet is forwarded along the faces indicated by the PIT entry, before the request is removed from the PIT.
    \item \emph{Forwarding Information Base} (FIB) aggregates the forwarding information at each host. Each FIB entry maps the content name to the outgoing faces and is associated at minimum with the tuple ``\{\emph{prefix, out-face(s)}\}". Received Data packets for pending Interests suggest a healthy content delivery path, hence the information associated with a successful Data packet delivery is used to update the FIB. Similar to the current Internet architecture, FIB in NDN also uses the longest prefix matching to select the outgoing faces.
\end{itemize}

Note that, current NDN proposal is based on routing at the higher levels in the name hierarchy in the network core, which even at the first level is expected to scale FIB to around 250 million entries and higher. For this paper, we assume routing at this top level hierarchy, particularly for the mobility study we assume routing at the level of application-ID. Considering this default routing setup, we next briefly explain the problems associated with mobility in information-centric networks and present our framework to study mobility in NDN.

\section{Mobility Framework}\label{Section:Mobility}

In information-centric networks, the problems associated with mobility are as follows:
\begin{itemize}
    \item \emph{Consumer mobility} causes Data packets to be returned to an inaccessible location (by the consumer). Furthermore, to receive these undelivered Data packets, \emph{Consumer} needs to initiate the recovery mechanism by retransmitting its Interest messages, with the network helping the recovery process through in-network caching\footnote{In-network caching, in general, limits Interest packet retransmissions by suppressing requests for the cached Data packets}. However, depending on the relative hop-by-hop distance between current and previous locations, and the shared hops along the path to destination, recovery process may introduce significant latency, enough to stall the consumer application, thereby causing unsatisfactory user experience.
    \item \emph{Producer mobility}, on the other hand, causes Interest messages to be forwarded towards the previous location (which essentially becomes an inaccessible location for the \emph{Producer} after the handover) until the Interest messages timeout. Depending on how the timeouts are handled at each router along the selected path (which is a recursive process along the downlink path towards the \emph{Consumer}), and how the \emph{Producer} announces its content availability to the network, we may also observe significant latency in the responses made by the \emph{Producer} during handovers.
\end{itemize}

In short, with host mobility, finding the content source can introduce significant latency to the system, making information-centric delivery a difficult choice, especially for time-sensitive applications (\emph{e.g.}, voice applications or live/interactive video streaming). If we also consider the possibility of vertical handovers\footnote{Vertical handovers typically refer to a host switching between, for instance, \emph{WiFi}, \emph{LTE}, or \emph{WiMAX} networks depending on bandwidth availability or signal quality.} (or a host moving across networks controlled by different entities or autonomous systems) during the activity lifetime of an application, problems associated with mobility will only get exacerbated as the path to the content may need to be re-established from scratch.

\begin{figure}
  \centering
  \includegraphics[clip,width=2.5in]{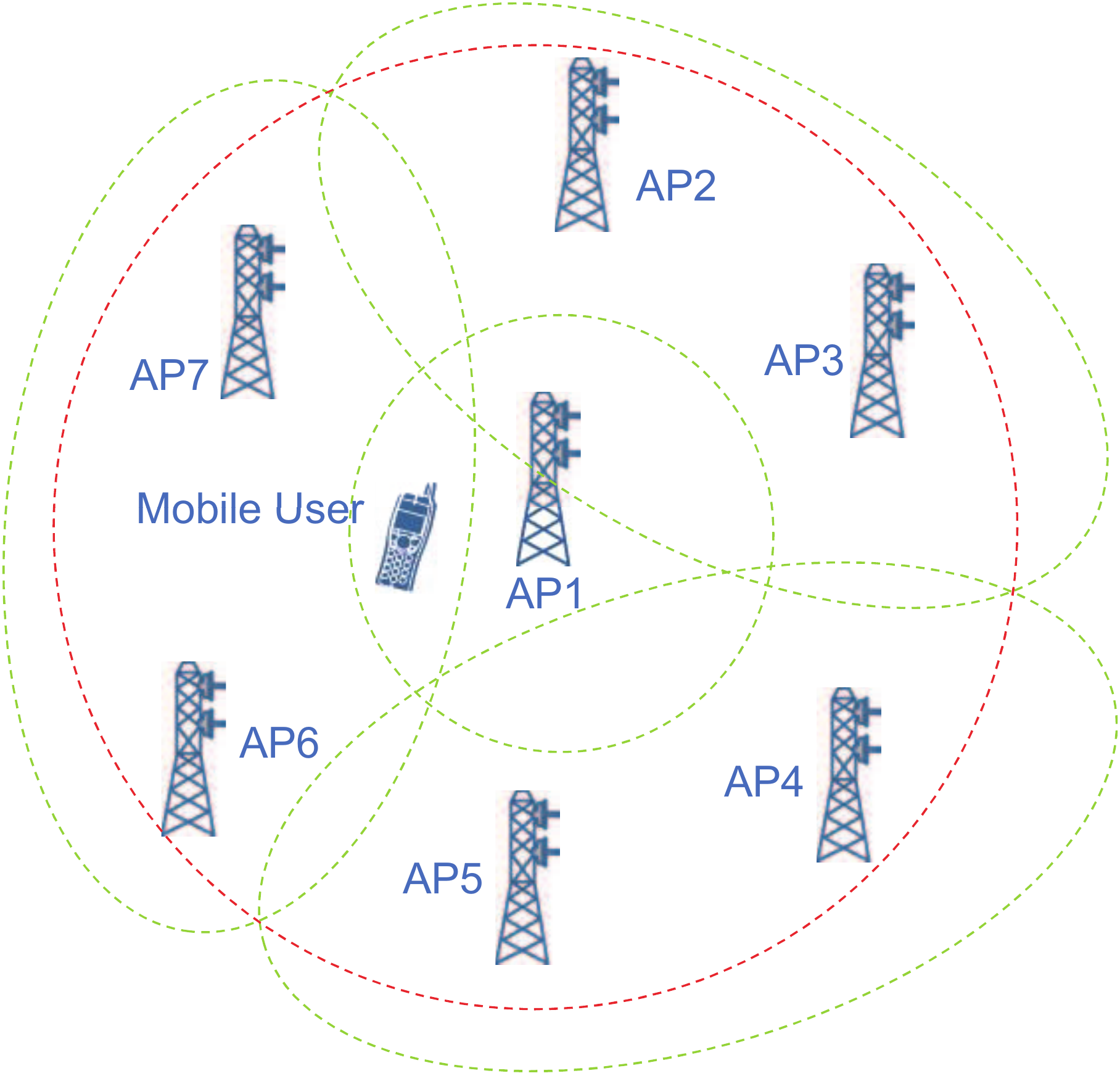}\\
  \caption{An example mobility scenario, with seven Point of Attachments (or Access Points) and a single mobile host.}\label{CCN_wireless_scenario_1}
\end{figure}

\begin{figure}[htb]
  \centering
  \includegraphics[clip,width=2.5in,height=1.8in]{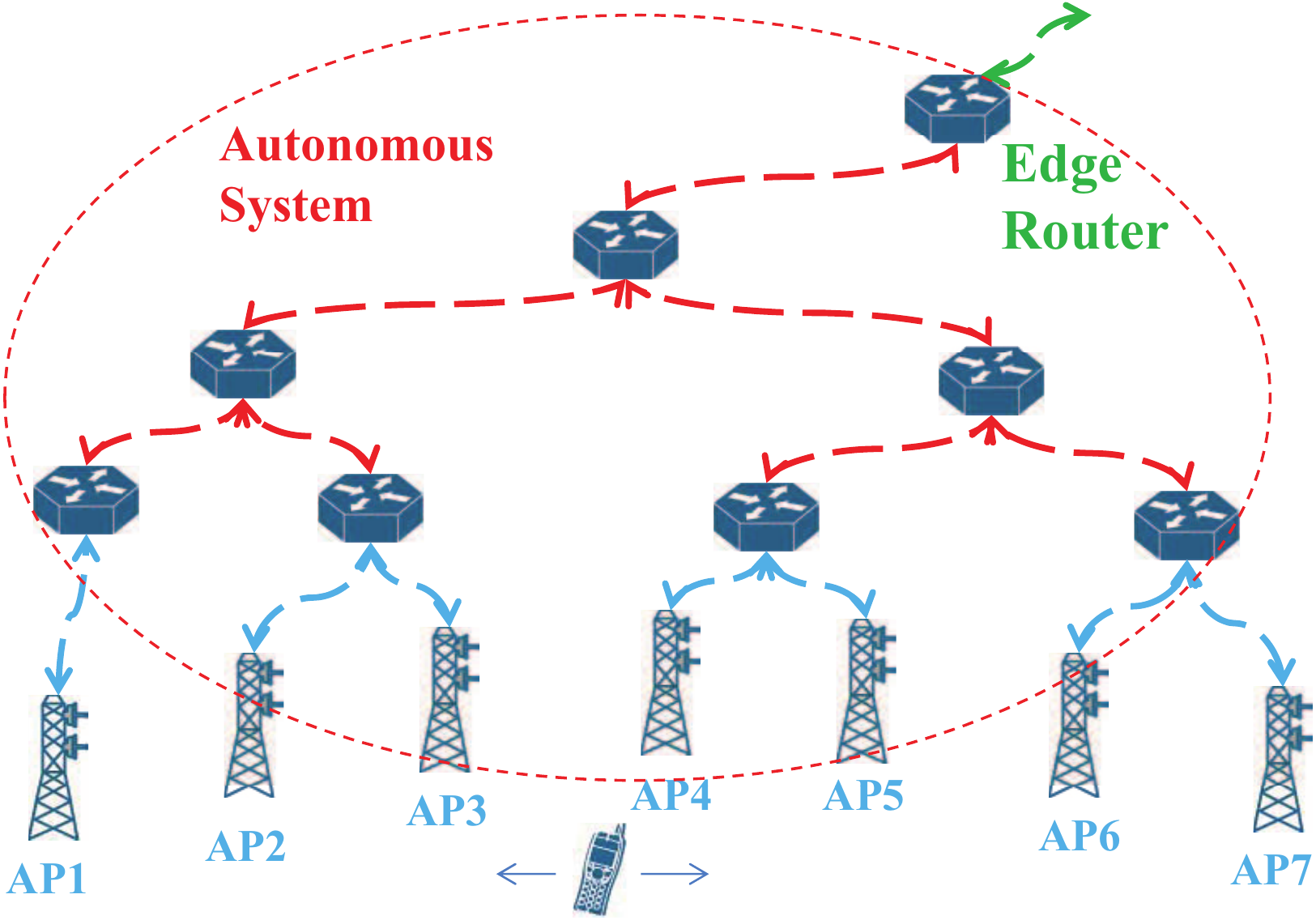}\\
  \caption{Backbone for the mobile network, single-AS scenario.}\label{CCN_wireless_scenario_2}
\end{figure}

To show the level of impact mobility can have on NDN, we focus on the wireless network topology shown in Figure~\ref{CCN_wireless_scenario_1}, which consists of seven access points within the mobile host's movement range. For the given topology, point of attachments (or access points) are connected hierarchically, towards the edge router of the host Autonomous System (AS). We show the backbone topology for the given network in Figure~\ref{CCN_wireless_scenario_2}.

\begin{figure}[htb]
  \centering
  \includegraphics[clip,width=3in,height=2in]{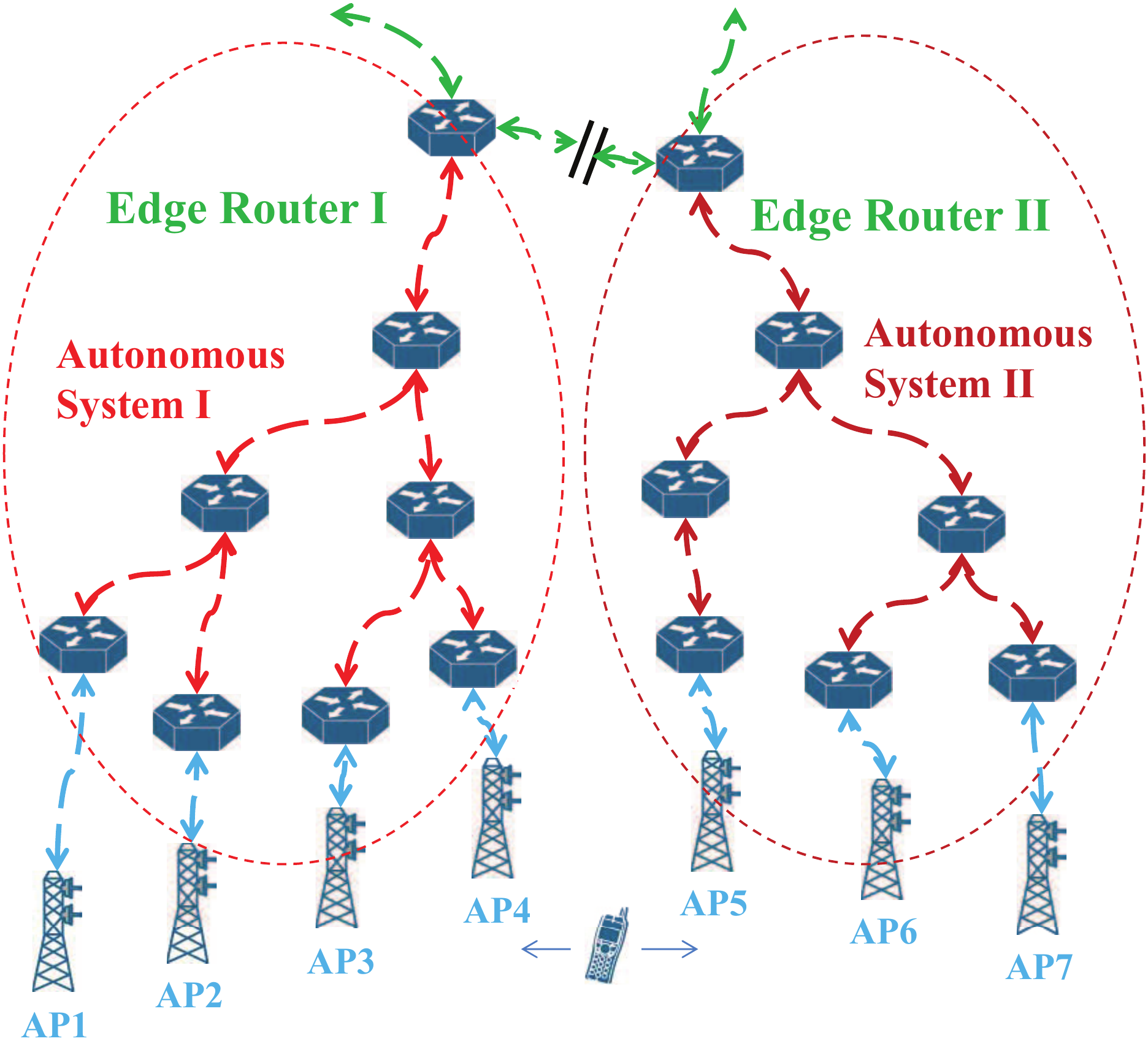}\\
  \caption{Backbone for the mobile network, 2-AS scenario.}\label{CCN_wireless_scenario_3}
\end{figure}

The baseline NDN scenario considered in our paper assumes the presence of two users in the network, a single \emph{Consumer} and a single \emph{Producer}, with each host placed in different ASs sharing the same hierarchical structure\footnote{Note that, proposed simulation framework allows us to specifically focus on mobility related problems by isolating the user traffic from other potential sources of performance degradation.}. We specifically focus on four scenarios in our study:
\begin{itemize}
\item \emph{Scenario 1}: one mobile host, one static host, each host is assigned a different AS (with the mobile host moving within a single AS), each of which observes the backbone structure illustrated in Figure~\ref{CCN_wireless_scenario_2} \footnote{\emph{Scenario I} allows us to observe the impact of consumer or producer mobility independently.},
\item \emph{Scenario 2}: Scenario 1 modified to allow for both hosts to be mobile,
\item \emph{Scenario 3}: one mobile host, one static host; each host is assigned two distinct ASs, with the mobile host moving within two ASs, for which the hierarchical backbone architecture is shown in Figure~\ref{CCN_wireless_scenario_3} \footnote{We used a bowtie topology at the core, \emph{i.e.}, we connected the edge routers (ERs) by directly connecting each consumer AS to each producer AS, with no direct connection between the two consumer/producer ASs. Specifically $ER_{AS1,con}\leftrightarrow\{ER_{AS3,pro},ER_{AS4,pro}\}$ and $ER_{AS2,con}\leftrightarrow\{ER_{AS3,pro},ER_{AS4,pro}\}$, where $\leftrightarrow$ represents the presence of a link.},
\item \emph{Scenario 4}: Scenario 3 modified to allow for both hosts to be mobile.
\end{itemize}

\section{Performance Analysis}\label{Section:Performance}

In our analysis, we focus on the performance of two core forwarding strategies in ndnSIM: $(i)$ \emph{Flooding}, which forwards Interest messages over all possible usable interfaces (referred to as green and yellow), and $(ii)$ \emph{Smart-flooding}, which floods Interest messages over all possible interfaces only if no green face is available\footnote{In ndnSIM, a third forwarding strategy is defined, which is referred to as the \emph{Best-route} strategy, where the possible interfaces are tried in order of priority, if no green face is available. We skipped its implementation here, since in a wireless setting, under the best-case scenario, its performance reduces to that of Smart-flooding. Also note that, we chose flooding-based strategies to limit the convergence latency associated with unicast based routing strategies due to host mobility.}. Additionally, we considered a third hybrid forwarding strategy, referred to as \emph{Semi-flooding}, which assumes \emph{Flooding} in the access network, and \emph{Smart-flooding} elsewhere\footnote{Semi-flooding can be considered as an optimized version of Smart-flooding with respect to handling the consumer mobility.}.

Since content delivery in information-centric networks is receiver-driven, \emph{Consumer}s pipeline Interests, even for unpublished (or \emph{yet-to-be-created}) content, to match the \emph{Producer} rate (\emph{e.g.} and maximize throughput by fully utilizing the bandwidth-delay product between \emph{Consumer} and \emph{Producer} \cite{VoCCN}).

\begin{figure*}
        \centering
        \hspace{-0.1in}
        \begin{subfigure}[b]{0.3\textwidth}\centering
               \includegraphics[clip,width=2.3in]{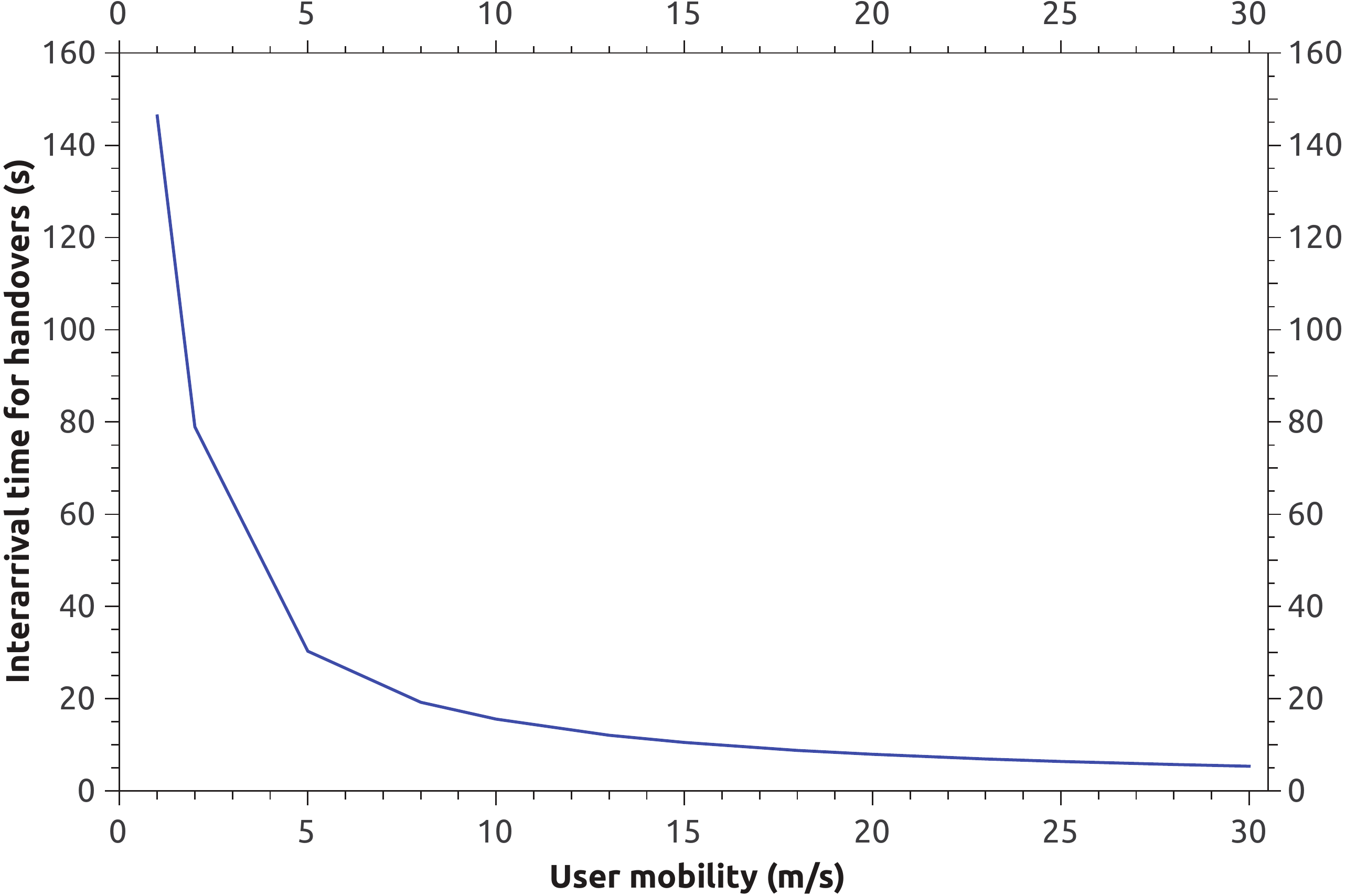}\\
               \caption{Mean interarrival time for handovers.}
               \label{AverageHandoverDuration}
        \end{subfigure}
        \hspace{0.1in}
        \begin{subfigure}[b]{0.3\textwidth}\centering
                \includegraphics[clip,width=2.3in,height=1.53in]{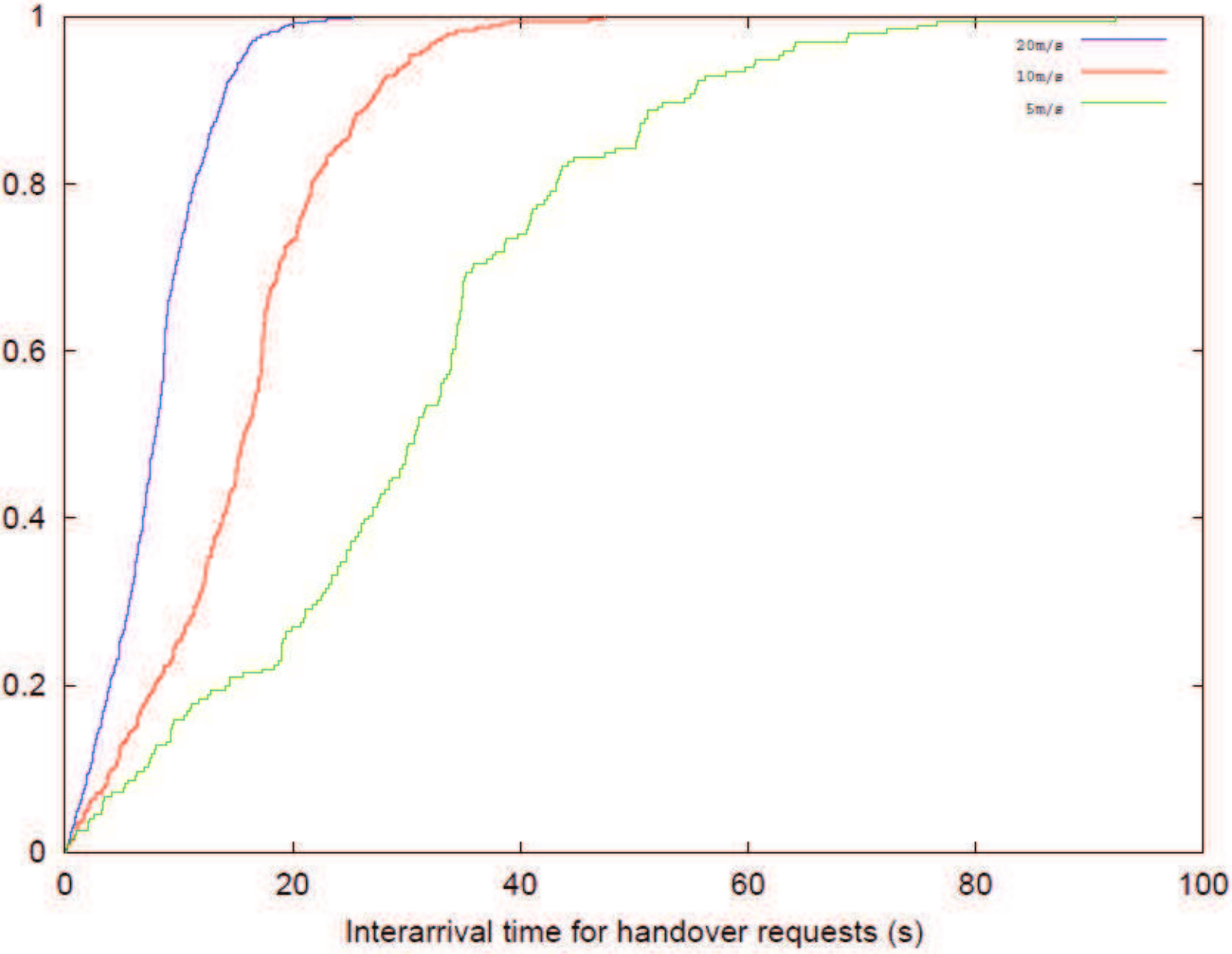}\\
                \caption{Empirical CDF for intra-AS handovers.}
                \label{HandoverDistribution}
        \end{subfigure}
        \hspace{0.1in}
        \begin{subfigure}[b]{0.3\textwidth}\centering
                \includegraphics[clip,width=2.3in,height=1.52in]{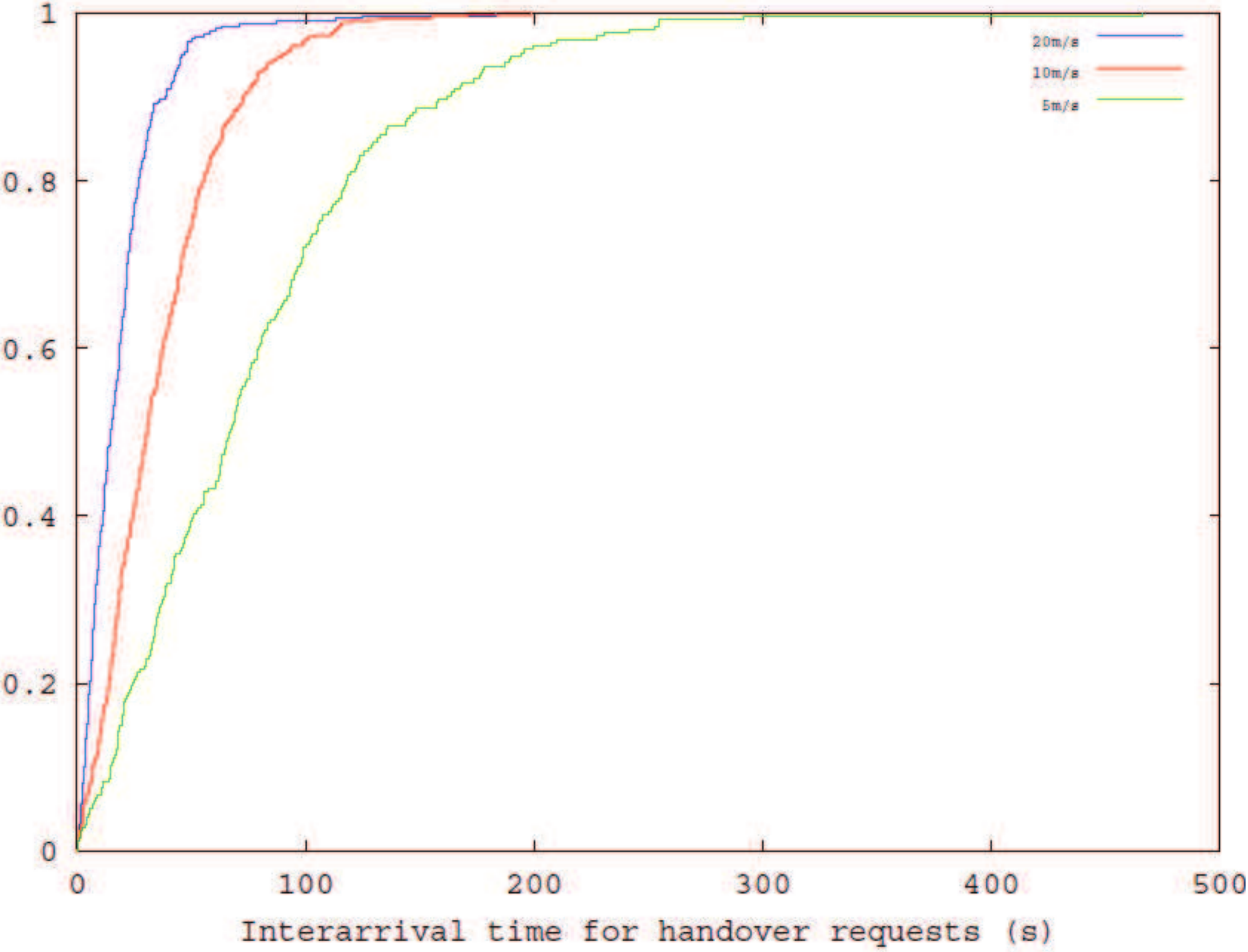}\\
                \caption{Empirical CDF for inter-AS handovers.}
                \label{HandoverDistribution2AS}
        \end{subfigure}
        \caption{Performance analysis for the handover requests at different host speeds.}\label{fig:handover_analysis}
\end{figure*}

\begin{figure*}
        \centering
        \hspace{-0.15in}
        \begin{subfigure}[b]{0.3\textwidth}\centering
               \includegraphics[clip,width=2.3in,height=1.5in]{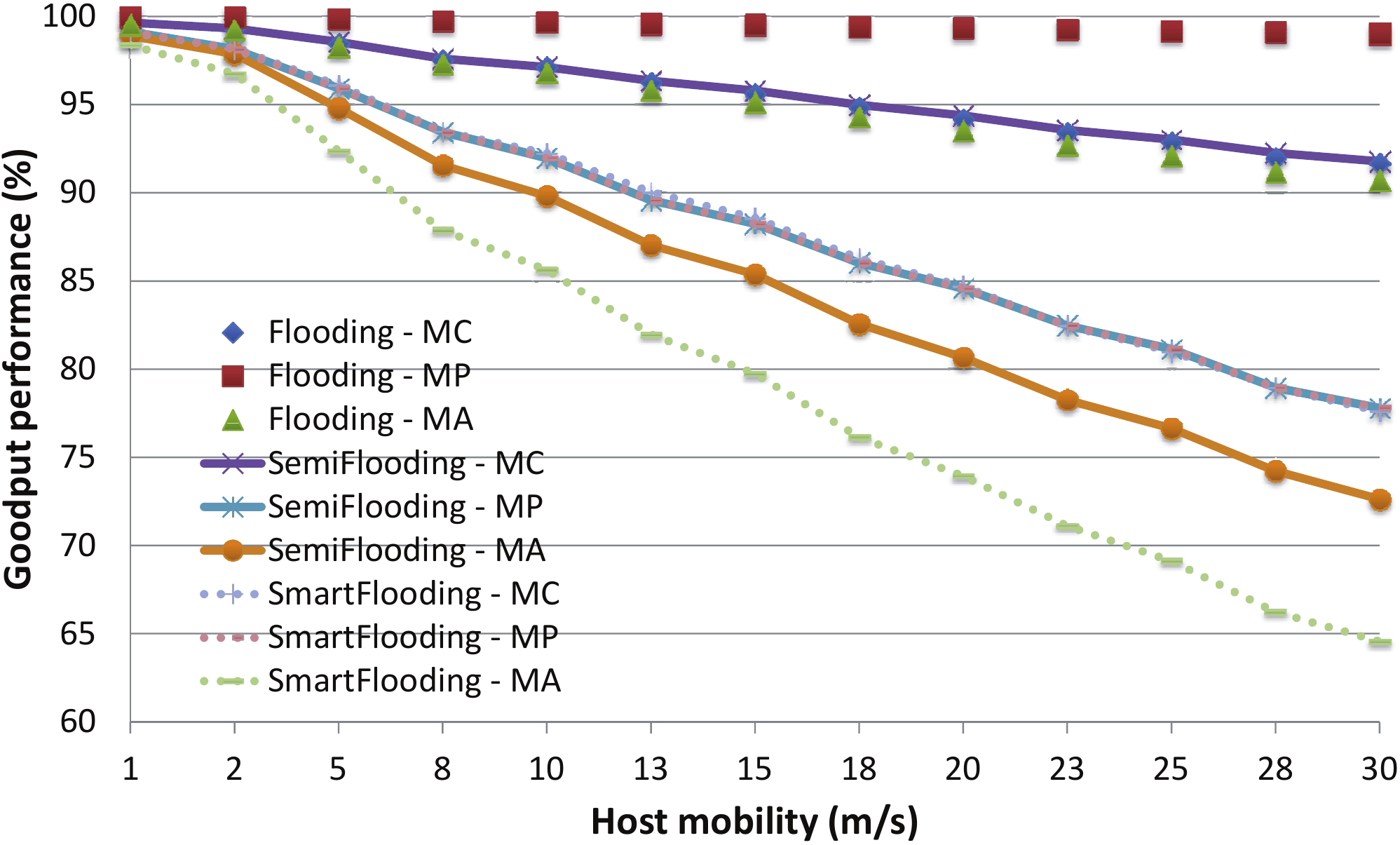}\\
               \caption{Uncorrelated data stream ($\delta = 500ms$).}
               \label{UncorrelatedGoodput}
        \end{subfigure}
        \hspace{0.05in}
        \begin{subfigure}[b]{0.3\textwidth}\centering
                \includegraphics[clip,width=2.3in,height=1.5in]{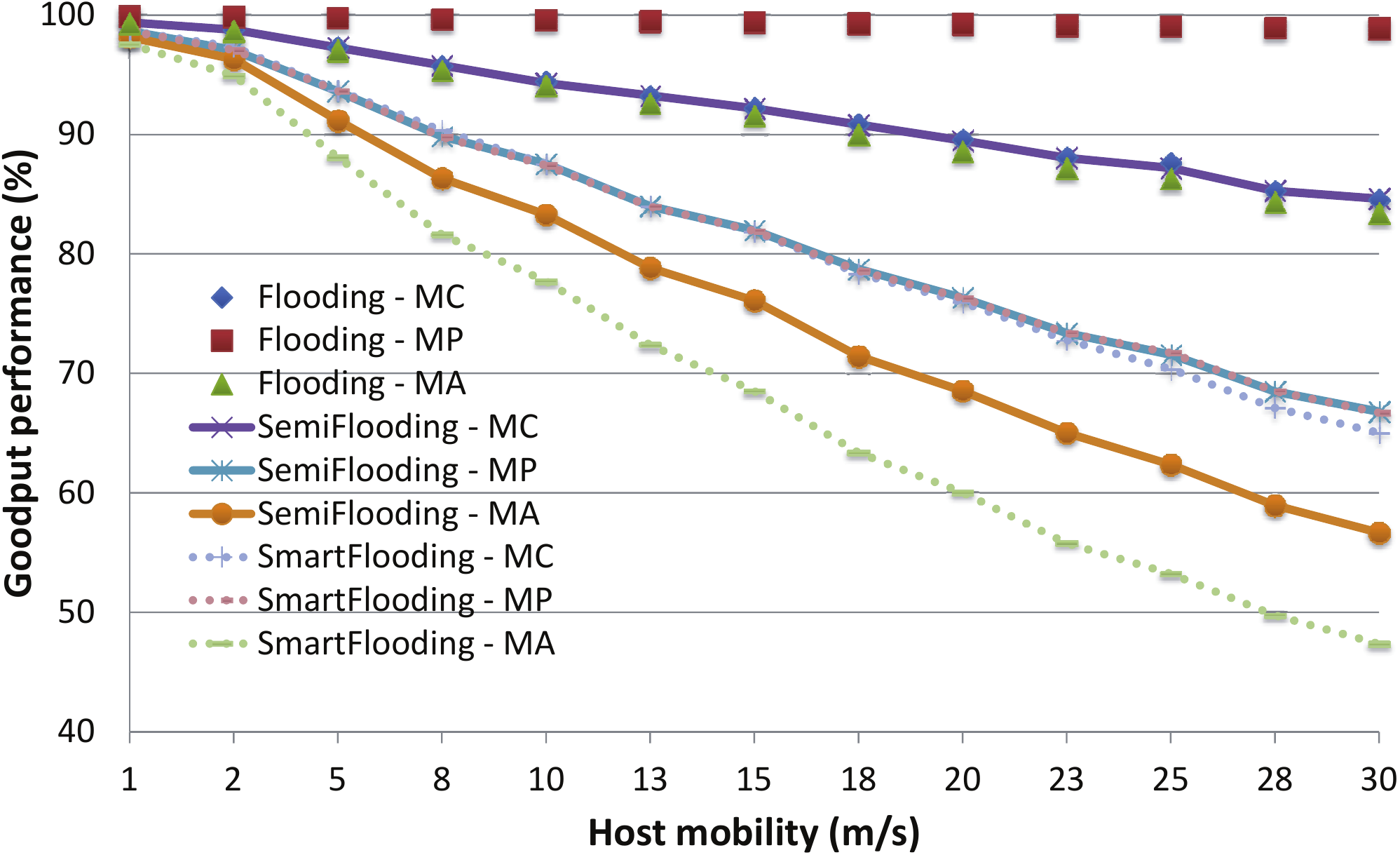}\\
                \caption{Correlated data stream ($\delta = 500ms$).}
                \label{CorrelatedGoodput250ms}
        \end{subfigure}
        \hspace{0.05in}
        \begin{subfigure}[b]{0.3\textwidth}\centering
                \includegraphics[clip,width=2.3in,height=1.5in]{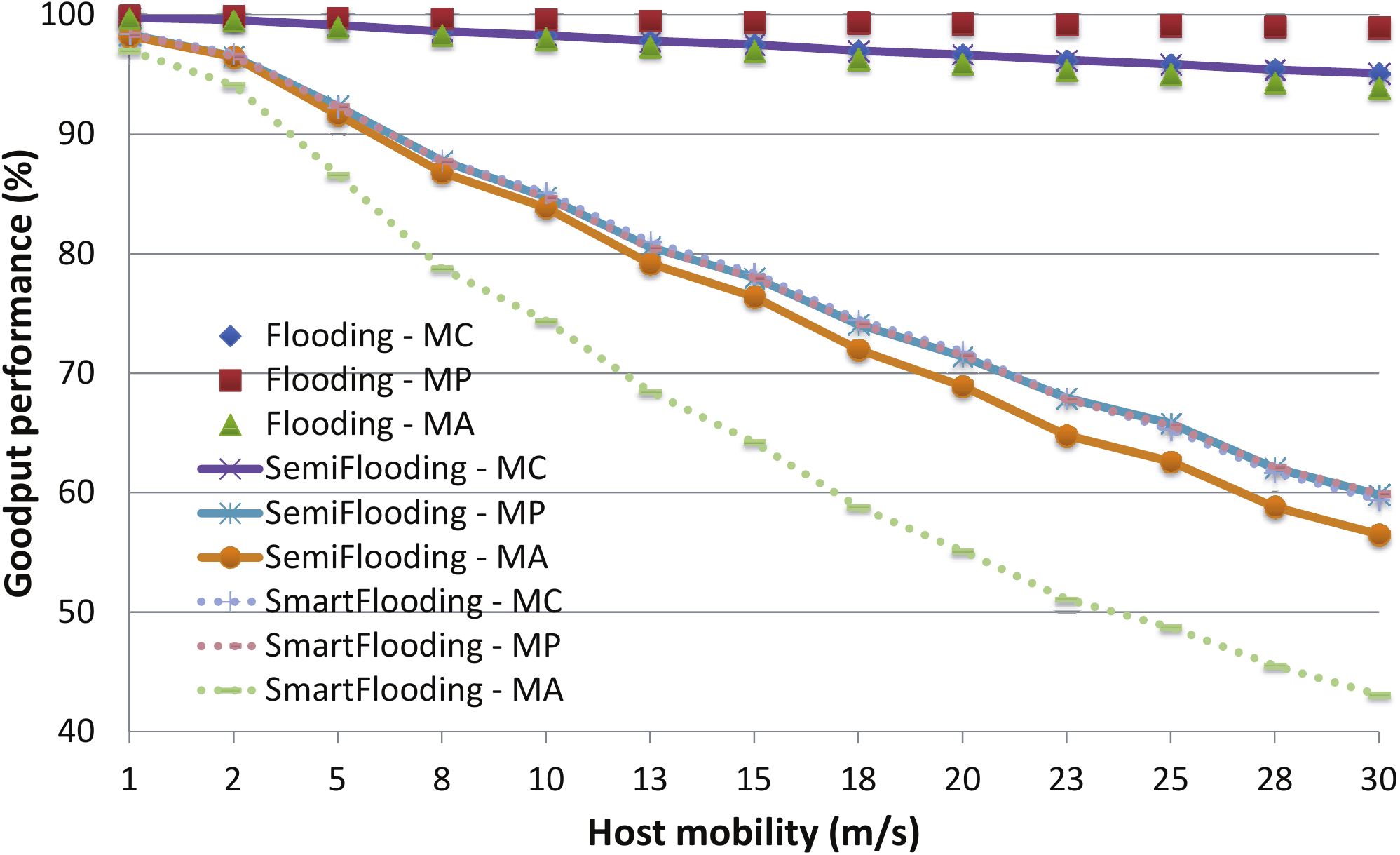}\\
                \caption{Correlated data stream ($\delta = 1s$).}
                \label{CorrelatedGoodput1000ms}
        \end{subfigure}
        \caption{User perceived Goodput performance for the delay-sensitive traffic with intra-AS mobility.}\label{fig:delay_sensitive_goodput}
\end{figure*}

In our study, we utilize the constant bit rate (CBR) traffic model, which acts as a good representative model for both the delay sensitive traffic (such as voice traffic) and the delay-tolerant traffic. In the case of delay-sensitive traffic, Interest messages are generated at the consumer side with constant interspacing between them. In our simulations, we assumed an Interest transmission rate of $50$ packets per second, which suggests an interspacing value of $20ms$ between consecutive Interest packet transmissions. Furthermore, we defined a default latency metric of $\delta=500ms$ for the delay-sensitive traffic to represent the associated servicing requirements and varied its value to show the impact of increased buffering.

We focused on two different scenarios for the delay-sensitive traffic: ($i$) \emph{uncorrelated} traffic flow (the impact of losing a packet is local), (ii) \emph{correlated} traffic flow (losing a packet can trigger additional losses\footnote{For instance, in the case of a video traffic, packet losses can lead to frames getting dropped.}). For the correlated traffic flow, we considered a synthetic 15fps video stream with a group-of-picture (GOP) duration of 1s\footnote{We tested NDN under different traffic scenarios, \emph{i.e.}, different frame rates and GOP structures, and observed similar trends in the observed performance. Due to space constraints, we only present the results corresponding to a single scenario.}. To show the results corresponding to the best case scenario, we assumed a GOP structure of IBBB (\emph{i.e.}, the impact of a frame loss propagates only if the frame that is lost is a key-frame). For the delay-tolerant traffic, we used a Poisson-based traffic generator (assuming the same request rate as the delay-sensitive traffic), and assumed no application-enforced latency constraint.

For the system parameters, we assume a transmission bandwidth of $5Mbps$ at each hop, and a hop-by-hop propagation latency of $10ms$. We used the default parameters for the wireless access network implementation\footnote{Specifically, we used the following setup for the wireless network: 802.11g operating at 11 Mbps(DSSS).} in ns3, and packet sizes of $28Bytes$, for the Interest message, and, $1024Bytes$, for the Data packet. We adjusted the distance between access points based on the perceived transmission range to ensure that a mobile host is always in-range of an AP within its movement region (\emph{i.e.}, distance between APs is set to $215m$ and each mobile host travels within a circular region that has a radius of $250m$, as illustrated with the outward circle in Figure~\ref{CCN_wireless_scenario_1}).

\subsection{Handover analysis}

Since handover process is not fully implemented in ndnSIM, we emulated the handover process by integrating each mobile host with multiple interface cards (each of which corresponds to an access point within the mobile host's movement region), and regularly updating the interfaces (\emph{i.e.}, enabling or disabling) depending on the perceived signal quality. A handover is initiated whenever ($i$) the signal quality perceived by the mobile host at the current network is not the best anymore, with respect to the signal quality perceived through at least one other access points within the host's range, {\emph{and}} $(ii)$ the respective change in signal quality persists for a specific duration. In our simulations we limited the handover duration to $50ms$\footnote{Note that, handover latency depends on many factors, including the wireless technology utilized by the host and the type of handover initiated by the mobile host. Varying its value will mostly affect the service quality of the interactive applications.}.


For the wireless network topology shown in Figure~\ref{CCN_wireless_scenario_1}, we varied the user mobility within $1-30m/s$.  To represent host mobility, we used the random waypoint model with zero pause time. The resultant handover performance is illustrated in Figure~\ref{AverageHandoverDuration}, which suggests, on average, a handover per $\approx 146.4ms$ for the lowest considered user speed, and a handover per $\approx 5.3ms$ for the highest considered user speed. We also illustrate the distribution for the handover request times in Figure~\ref{HandoverDistribution} for three different user mobility scenarios: $\{5m/s,10m/s,20m/s\}$. For each of these cases, we observe that the distributions for the handover requests closely follow a linear trend suggesting uniformly distributed interarrival times.


For the inter-AS handovers, we observe a longer-tailed distribution for which the results corresponding to three different mobility scenarios are shown in Figure~\ref{HandoverDistribution2AS}. On average, we observed that $\approx 40\%$ of the handover requests correspond to inter-AS handovers.


\subsection{Intra-AS mobility}


\begin{figure*}
        \centering
        \hspace{-0.15in}
          \begin{subfigure}[b]{0.3\textwidth}\centering
                \includegraphics[clip,height=1.5in,width=2.3in]{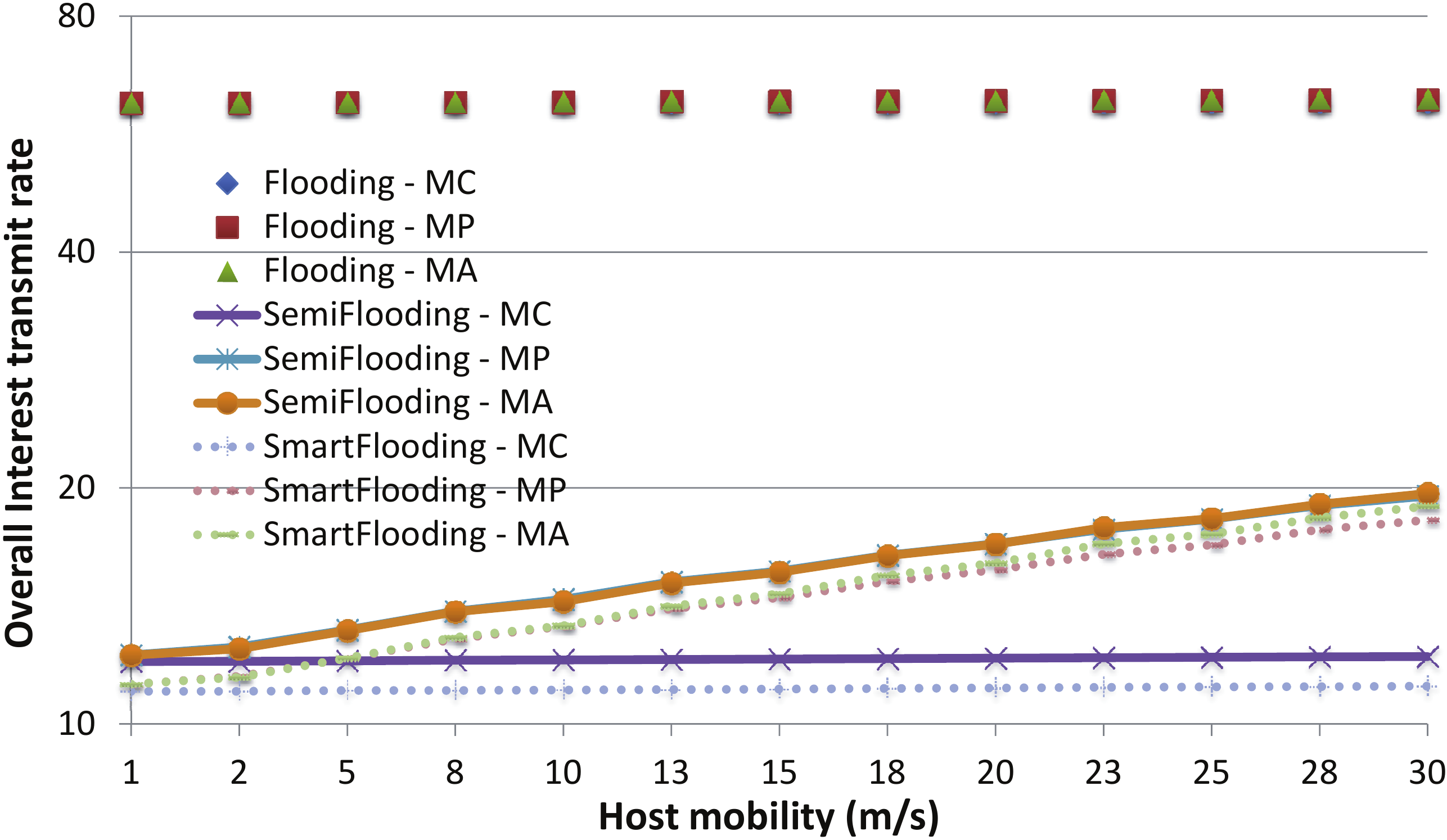}\\
                \caption{Delay-tolerant traffic.}
                \label{DataOverhead}
        \end{subfigure}
        \hspace{0.05in}
        \begin{subfigure}[b]{0.3\textwidth}\centering
               \includegraphics[clip,width=2.3in,height=1.5in]{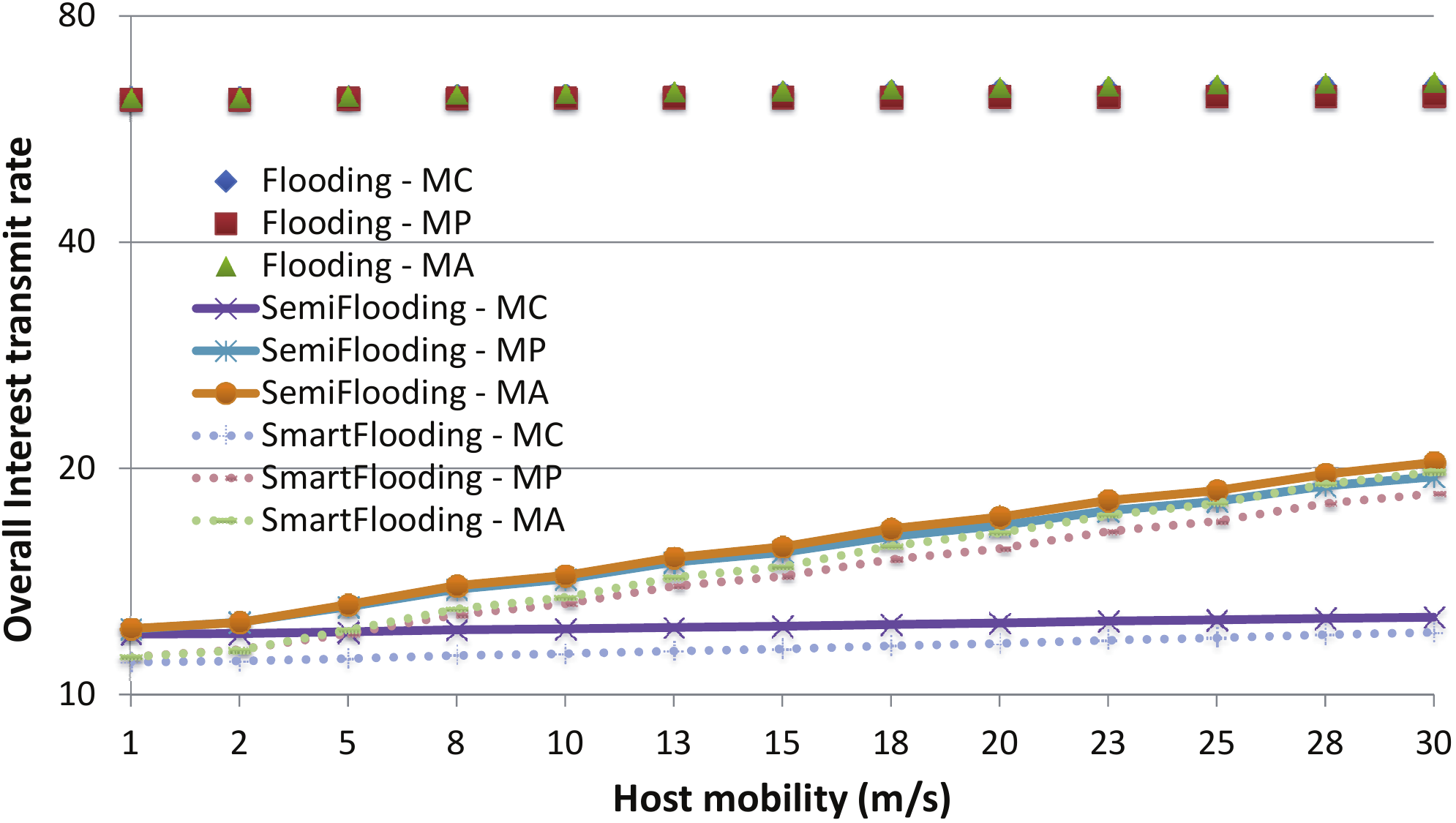}\\
               \caption{Delay-sensitive uncorrelated data stream.}
               \label{UncorrelatedOverhead}
        \end{subfigure}
        \hspace{0.05in}
        \begin{subfigure}[b]{0.3\textwidth}\centering
                \includegraphics[clip,width=2.3in,height=1.5in]{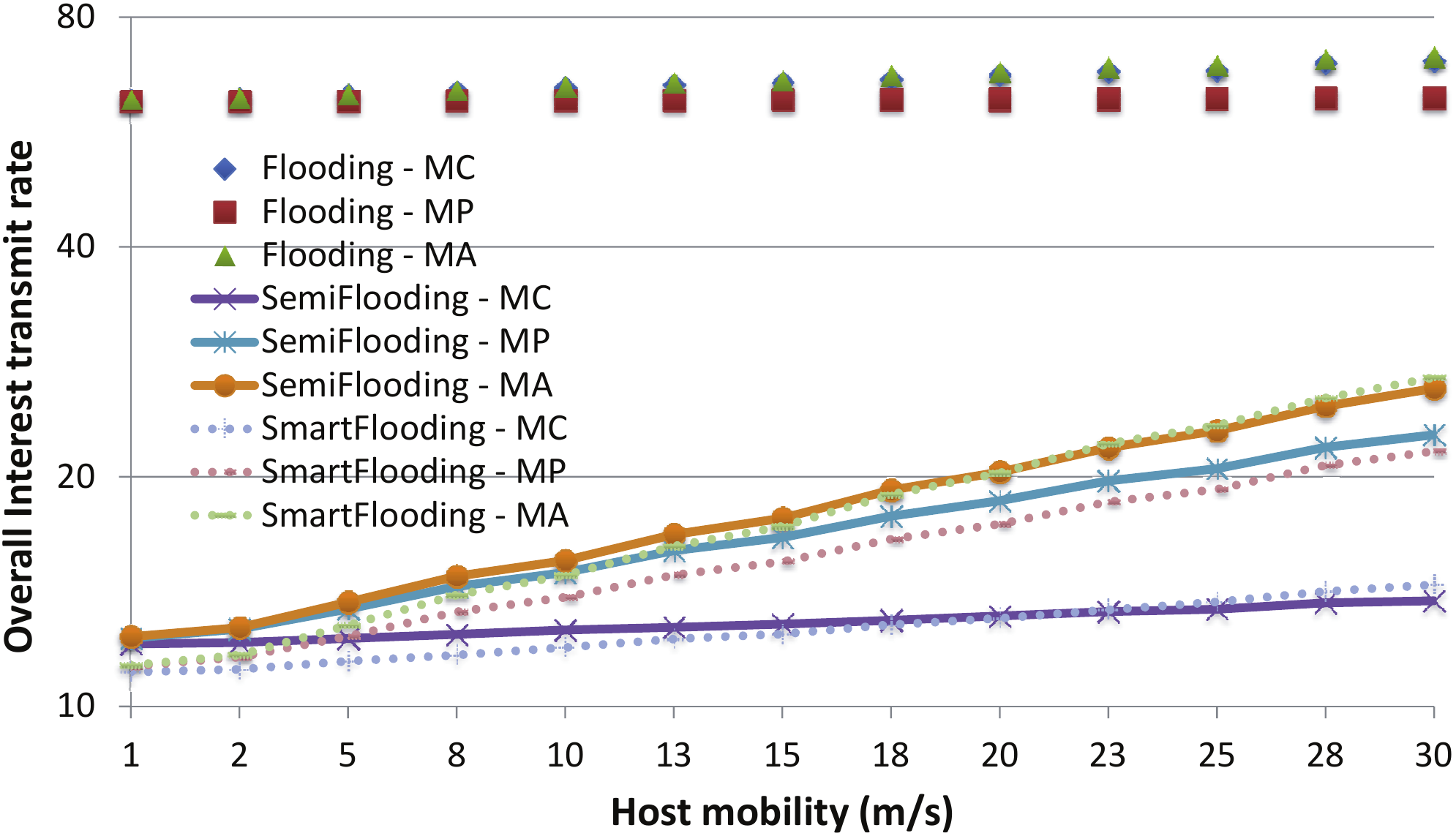}\\
                \caption{Delay-sensitive correlated data stream.}
                \label{CorrelatedOverhead}
        \end{subfigure}
        \caption{Average number of Interest messages transmitted per successfully received Data packet with intra-AS mobility ($\delta=500ms$ for delay-sensitive scenario).}\label{fig:overhead}
\end{figure*}

We show the comparative results for the Goodput performance (\emph{i.e.}, effective throughput for delay-sensitive traffic) in Figure~\ref{fig:delay_sensitive_goodput}, where $\{MC,MP,MA\}$ represent \{\emph{mobile-consumer, mobile-producer, mobile-all}\}\footnote{In our analysis we use the term \emph{Goodput} to represent the performance of delay-sensitive traffic, and \emph{Throughput} to represent the performance of delay-tolerant traffic.}. We observe that Flooding, as expectedly, achieves the best performance even when the hosts are highly mobile (\emph{e.g.}, performance loss is limited to $\approx 10-20\%$ at the highest mobility level), but at the cost of higher overhead as we explain shortly. In contrast, we observe significant performance degradations with both Smart-flooding and Semi-flooding (even though Semi-flooding achieves overall better results), up to $\approx 42\%$ performance loss when either one of the hosts is mobile, and up to $\approx 56\%$ performance loss when both hosts are mobile. We also observe that only Flooding takes advantage of the increase in start-up latency, as both Smart-flooding and Semi-flooding allocate more of their bandwidth to retransmissions.

Furthermore, we observe the \emph{mobile-consumer} scenario to be performing worse than the \emph{mobile-producer} scenario with Flooding, whereas the opposite is observed with Smart-flooding and Semi-flooding. The difference is typically caused by how the timeouts and recoveries are handled within the network. Flooding re-establishes the connection proactively, once the \emph{mobile-producer} attaches to an access point, whereas, without continuous flooding, \emph{Consumer} or a node along the previously-active path between \emph{Consumer} and \emph{Producer} needs to re-initiate the connection reactively, using various triggers (\emph{i.e.}, NACK or timeout). 

\begin{figure}[htb]
  \centering
  \includegraphics[clip,width=2.4in]{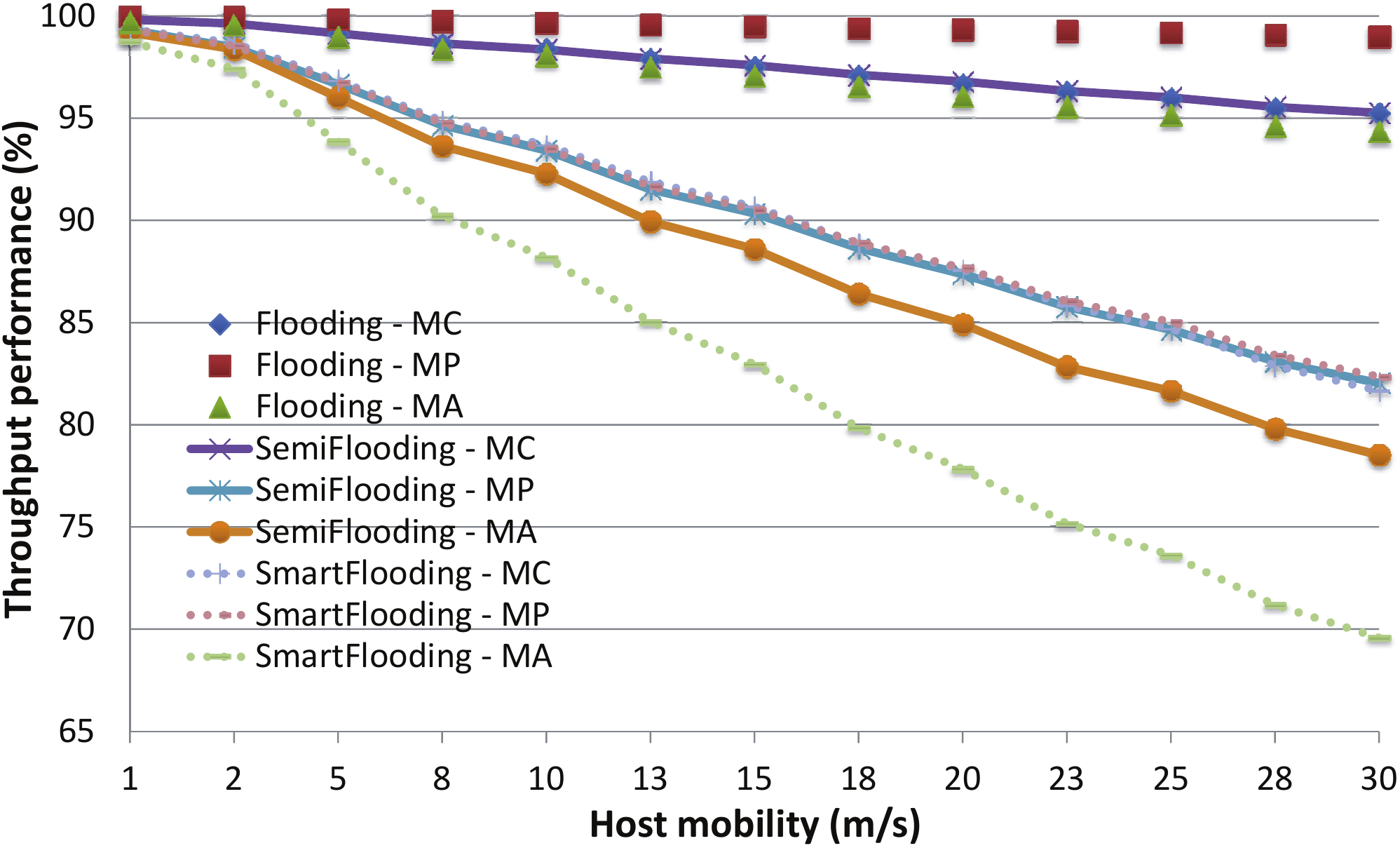}\\
  \caption{Throughput performance comparison between Flooding, Smart-flooding, and Semi-Flooding, for delay-tolerant traffic, with intra-AS mobility.}\label{ThroughputResults}
\end{figure}

We next present the results for the Throughput in Figure~\ref{ThroughputResults}, for delay-tolerant traffic. Similar to the Goodput results, we continue to observe significant performance losses, up to $\approx 18\%$ for the single mobile-host scenario and up to $\approx 31\%$ for the all-mobile scenario, which is due to timeouts caused by the handover events, and the retransmissions involved during the delivery of Interest/Data packets after timeouts. The above results also suggest that, for delay-tolerant traffic, in-network caches offer higher performance gains, when compared to delay-sensitive traffic.


Next, we show the results for the messaging overhead in Figure~\ref{fig:overhead}. We observe that delay-tolerant traffic and uncorrelated delay-sensitive traffic observe similar trends for the effective Interest transmission rate (with slight increase for the delay-sensitive scenario), which is caused by the increased recovery rate (due to better application support for retransmissions). On the other hand, for correlated delay-sensitive traffic, we observe a noticeably worsened performance (up to $15\%$ for Flooding, up to $45\%$ for Smart-flooding and Semi-flooding). Furthermore, except for the Flooding strategy (which introduces a nearly constant overhead, regardless of the user speeds or the application type), as the mobility increases, we observe an increase in overhead, as the user fails to recover from its losses at the desired rate. These results have to be compared to the static-hosts case where for the given setup results in, on average, $11$ Interest message transmissions per successfully received/decoded Data packet. Hence, with Smart-flooding and Semi-flooding strategies, for the given network topology, we observe up to $2-2.5$ times increase in messaging overhead, whereas with the Flooding strategy, that rate increases to $6-7$. Also note that, Flooding strategy creates multiple data paths between \emph{Consumer} and \emph{Producer}, resulting in an even larger overhead.


Additionally, we observe that \emph{Producer} mobility introduces higher overhead, when compared to \emph{Consumer} mobility. The reason for that is, in the case of \emph{Consumer} mobility, Interest flooding always starts at the \emph{Consumer} side as it is the only host that is affected by the timeouts. \emph{Producer} mobility, on the other hand, involves a higher number of NDN routers that can initiate the recovery process, hence cause an increased number of recovery attempts. We can mitigate the \emph{Producer} mobility problem with the help of NACKs, but we are still limited in the number of attempts that we may need to perform (\emph{i.e.}, lower-bounded by the \emph{mobile-consumer} results) to successfully deliver the Interest messages to the \emph{Producer} or an in-network cache.

\subsection{Inter-AS mobility}
\begin{figure*}
        \centering
        \hspace{-0.15in}
        \begin{subfigure}[b]{0.3\textwidth}\centering
               \includegraphics[clip,width=2.3in,height=1.5in]{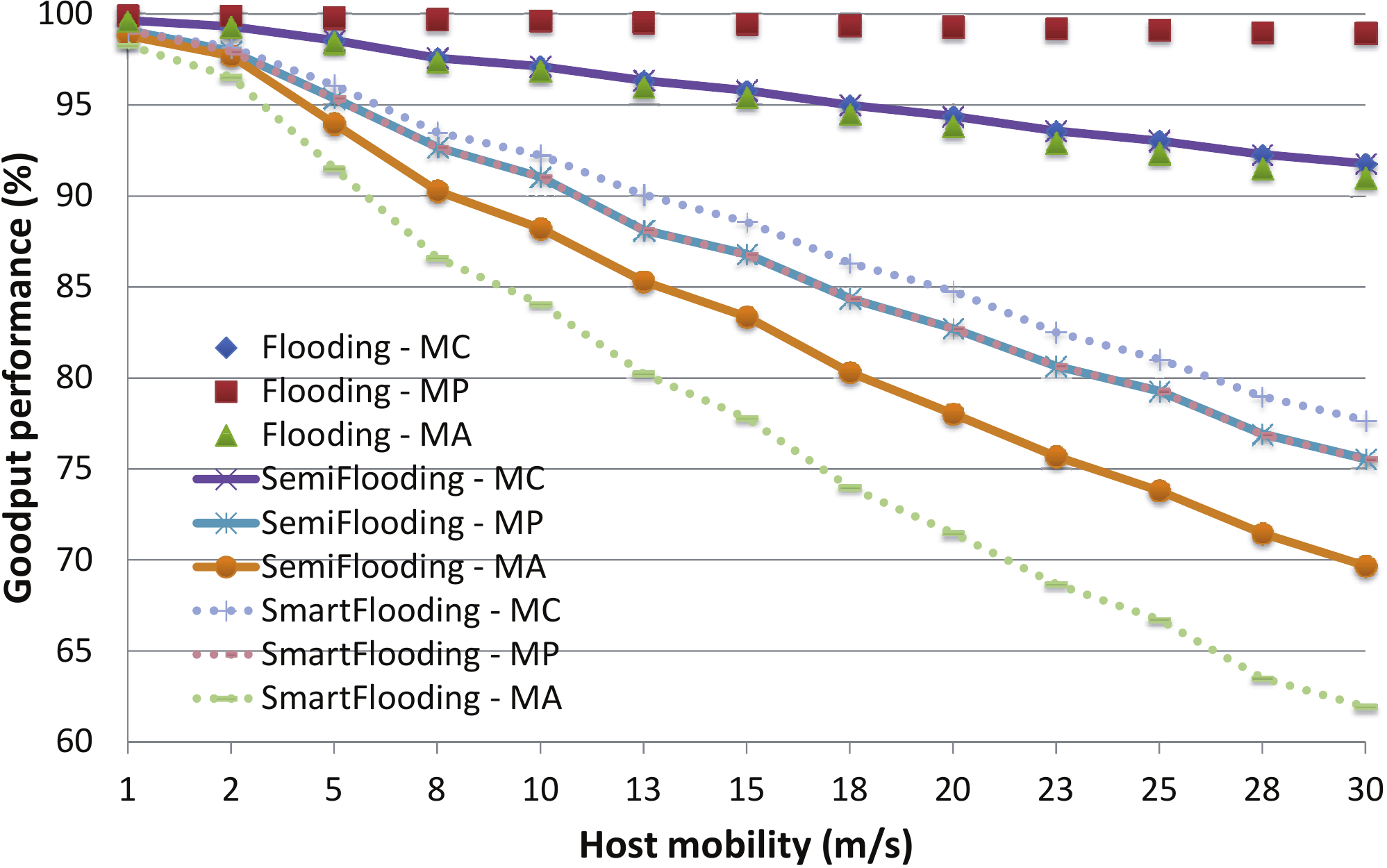}\\
               \caption{Uncorrelated data stream ($\delta = 500ms$).}
               \label{2ASUncorrelatedGoodput}
        \end{subfigure}
        \hspace{0.05in}
        \begin{subfigure}[b]{0.3\textwidth}\centering
                \includegraphics[clip,width=2.3in,height=1.5in]{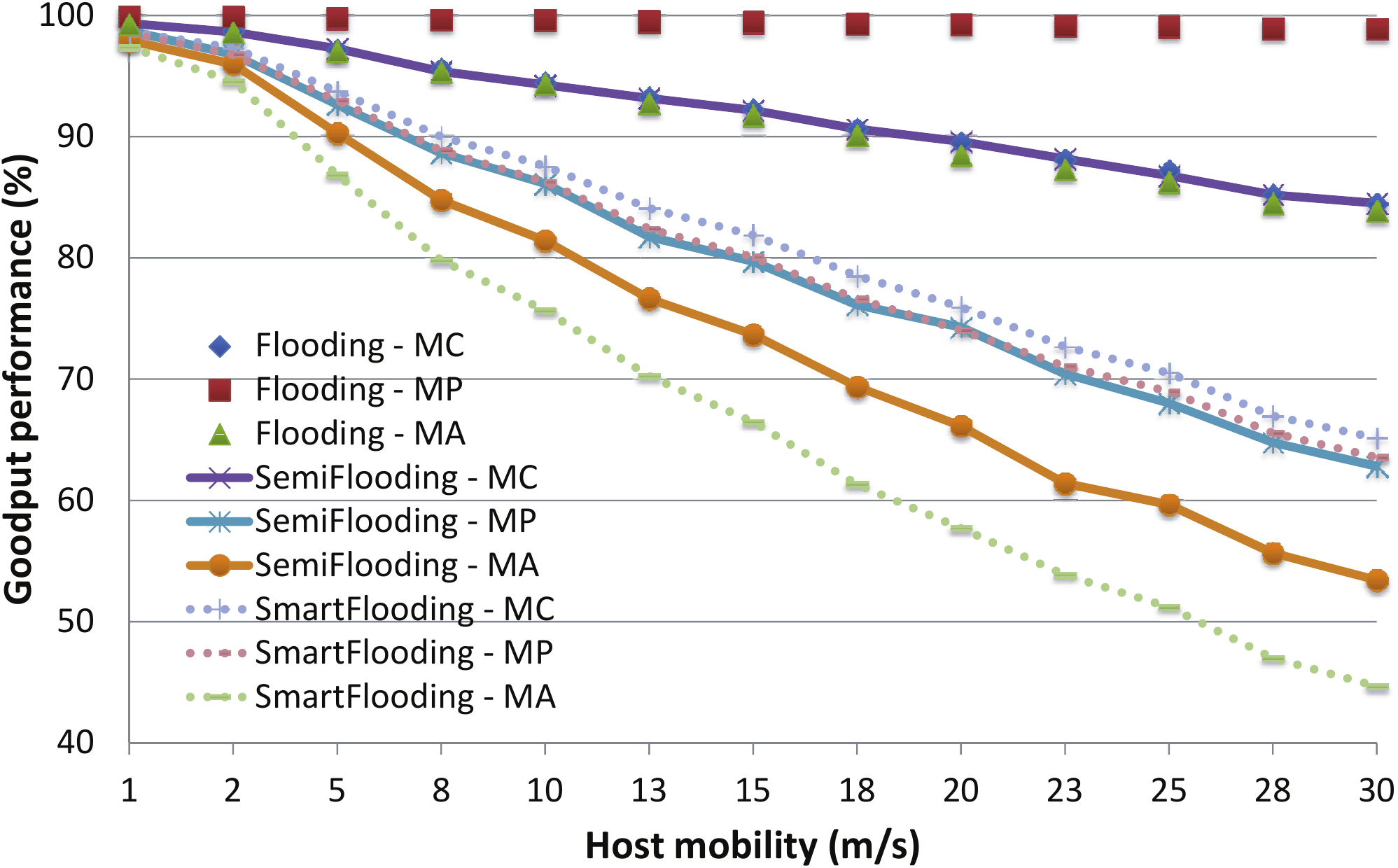}\\
                \caption{Correlated data stream ($\delta = 500ms$).}
                \label{2ASCorrelatedGoodput250ms}
        \end{subfigure}
        \hspace{0.05in}
        \begin{subfigure}[b]{0.3\textwidth}\centering
                \includegraphics[clip,width=2.3in,height=1.5in]{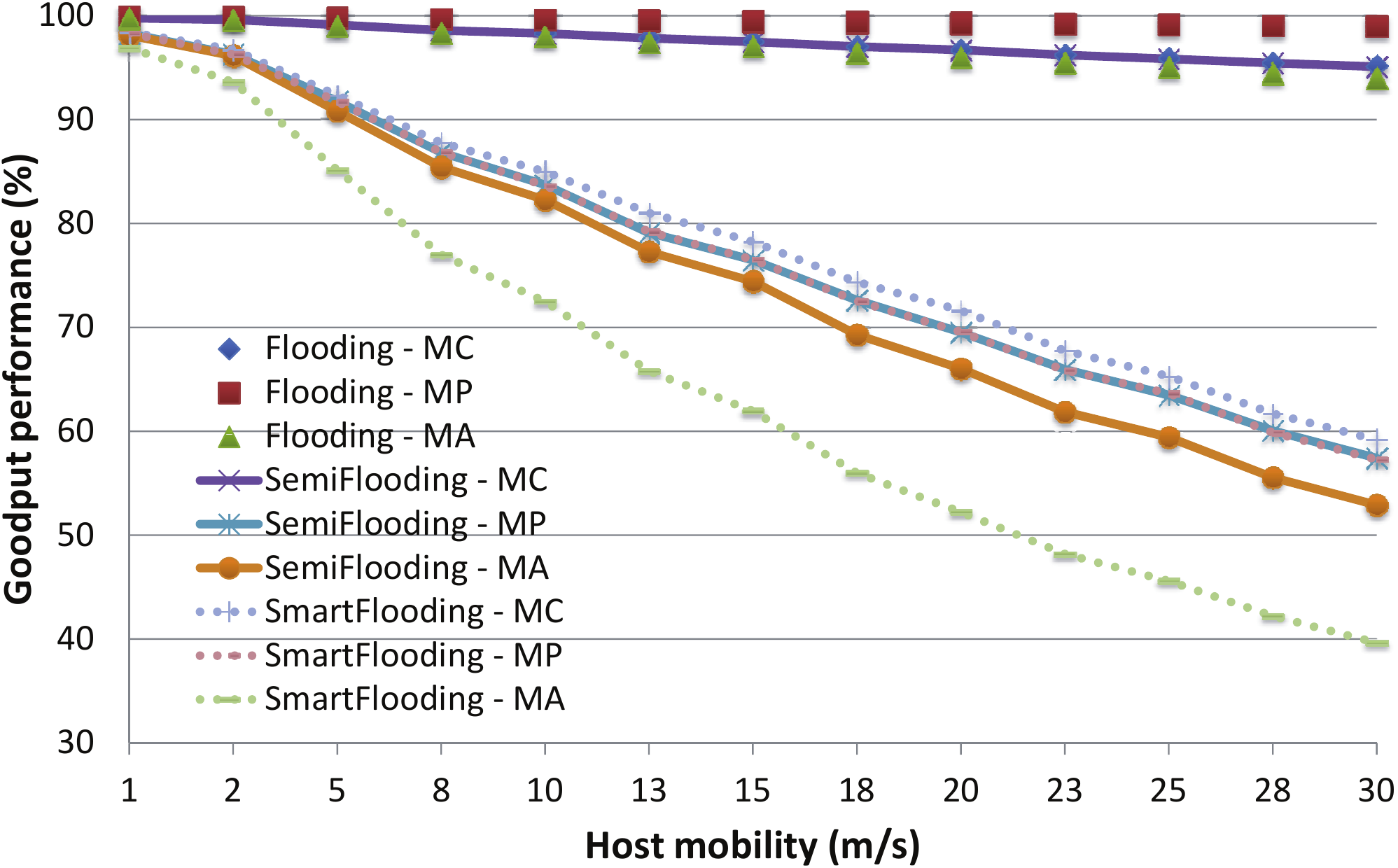}\\
                \caption{Correlated data stream ($\delta = 1s$).}
                \label{2ASCorrelatedGoodput1000ms}
        \end{subfigure}
        \caption{User perceived Goodput performance for the delay-sensitive traffic with inter-AS mobility.}\label{fig:delay_sensitive_goodput_2as}
\end{figure*}

\begin{figure}[htb]
  \centering
  \includegraphics[clip,width=2.4in]{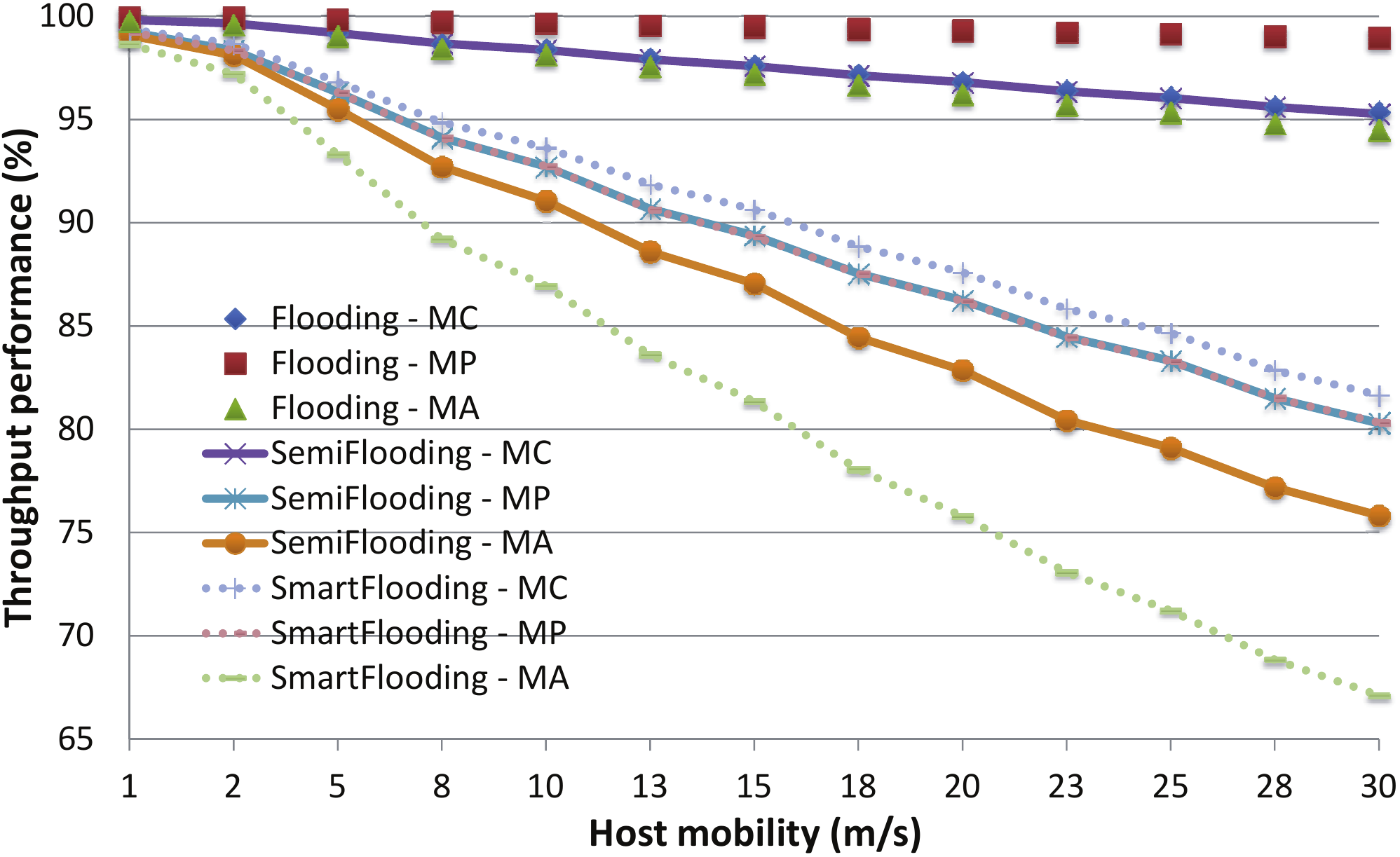}\\
  \caption{Throughput performance comparison between Flooding, Smart-flooding, and Semi-Flooding, for delay-tolerant traffic, with inter-AS mobility.}\label{ThroughputResults2AS}
\end{figure}

We next investigate the impact of mobility among different autonomous systems. In our simulations, we limit the distance between two ASs to two hops. We show the comparative \emph{Goodput} and \emph{Throughput} results in Figures~\ref{fig:delay_sensitive_goodput_2as} and \ref{ThroughputResults2AS}, respectively. We observe that the \emph{mobile-consumer} case is not affected by the inter-AS mobility, whereas introducing a \emph{mobile-producer}, we observed a noticeable drop in performance ($\approx 5-10\%$ decrease). Even though the \emph{mobile-producer} case involves an increased number of attempts during recovery (through Interest flooding) after the timeouts, inter-AS mobility necessitates the Interest messages to go through nearly double the number of hops when compared to the \emph{mobile-consumer} case. As a consequence of increased recovery latency and increased flooding rates, we observe performance degradations for both types of traffic in the \emph{mobile-producer} case.

%
%

\begin{figure*}
        \centering
        \hspace{-0.15in}
          \begin{subfigure}[b]{0.3\textwidth}\centering
                \includegraphics[clip,width=2.3in,height=1.5in]{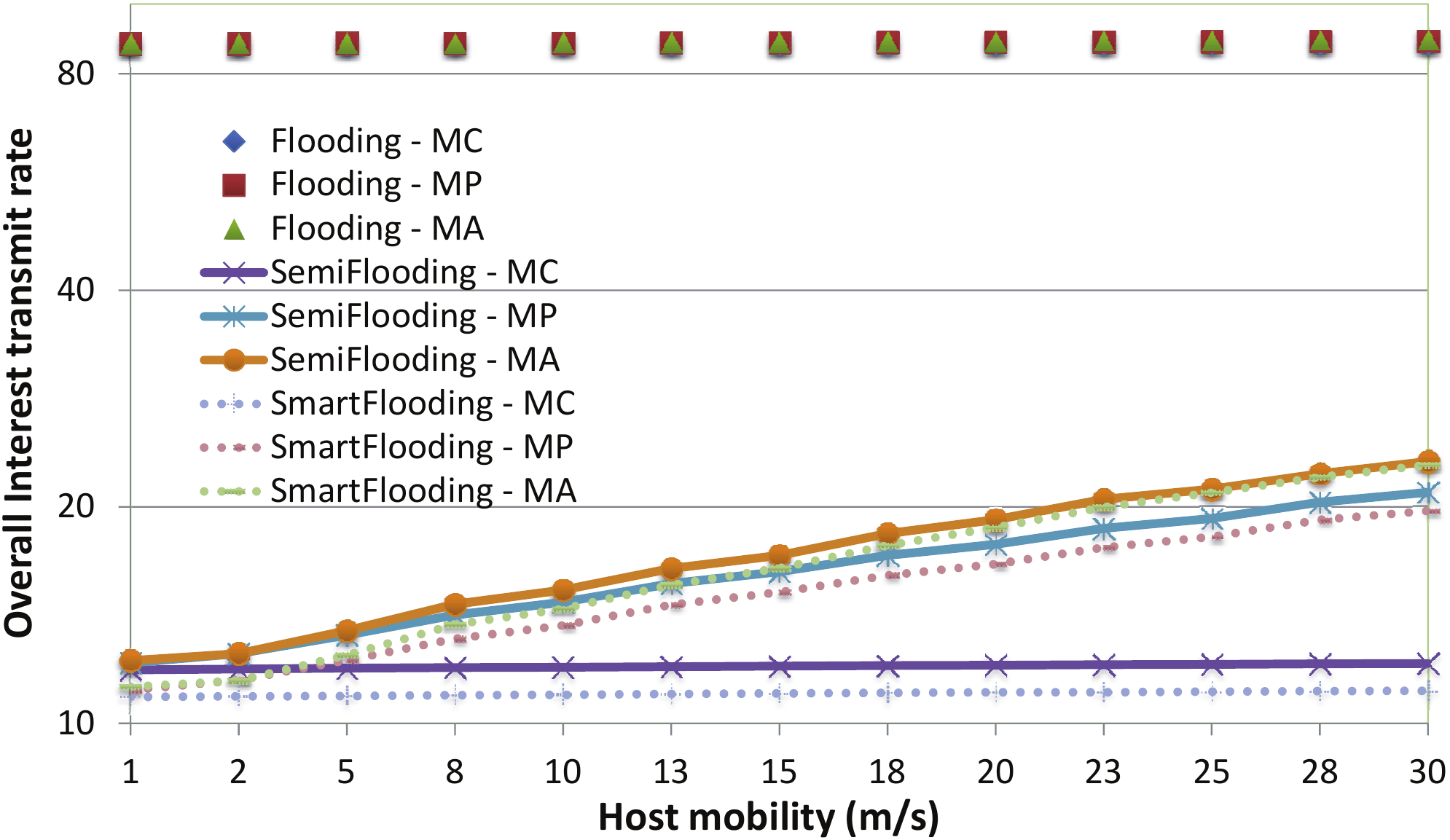}\\
                \caption{Delay-tolerant traffic.}
                \label{DataOverhead2AS}
        \end{subfigure}
        \hspace{0.05in}
        \begin{subfigure}[b]{0.3\textwidth}\centering
               \includegraphics[clip,width=2.3in,height=1.5in]{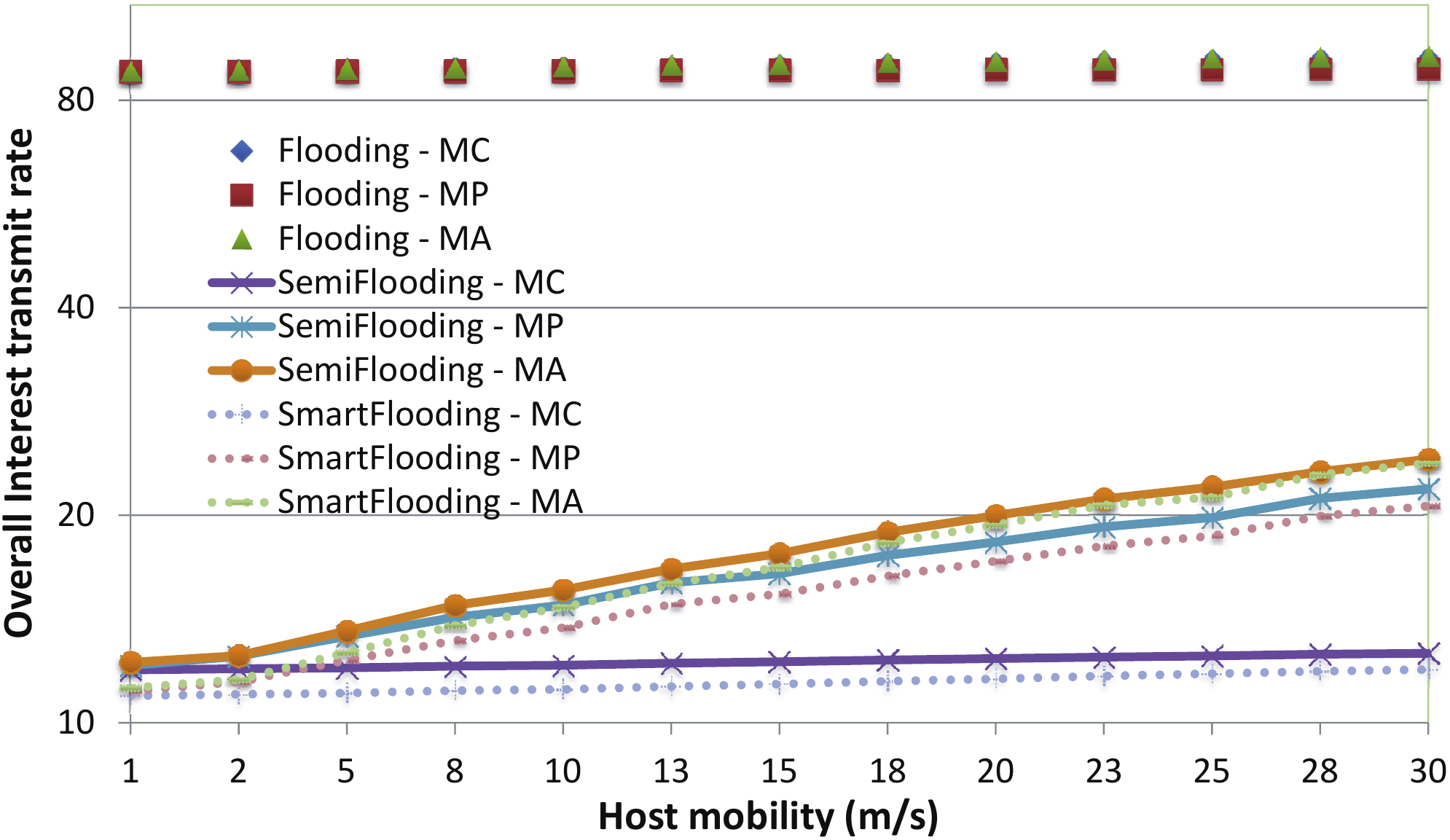}\\
               \caption{Delay-sensitive uncorrelated data stream.}
               \label{UncorrelatedOverhead2AS}
        \end{subfigure}
        \hspace{0.05in}
        \begin{subfigure}[b]{0.3\textwidth}\centering
                \includegraphics[clip,width=2.3in,height=1.5in]{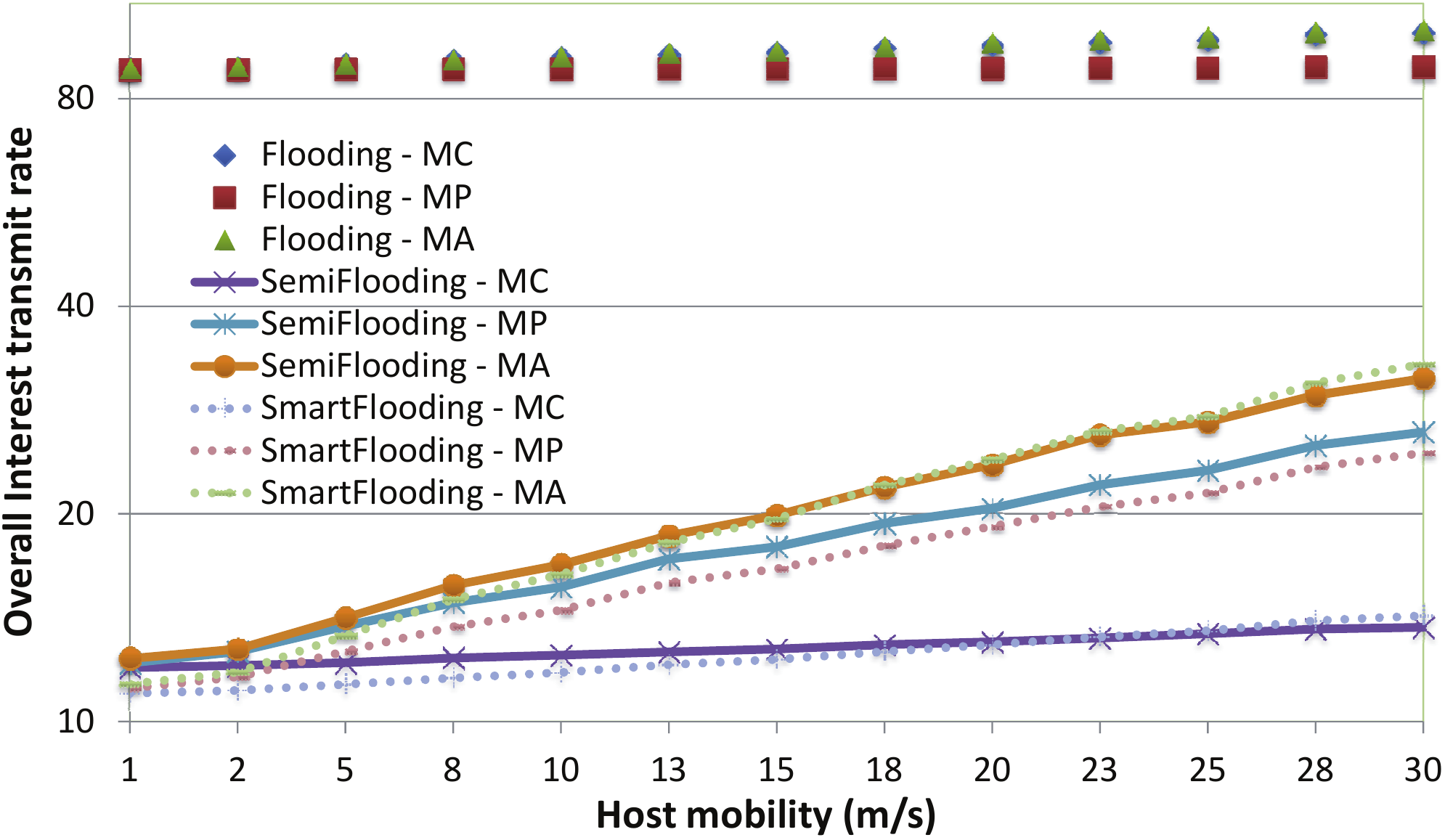}\\
                \caption{Delay-sensitive correlated data stream.}
                \label{CorrelatedOverhead2AS}
        \end{subfigure}
        \caption{Average number of Interest messages transmitted per successfully received Data packet with inter-AS mobility ($\delta=500ms$ for delay-sensitive scenario).}\label{fig:overhead_2AS}
\end{figure*}

We can also see the impact of inter-AS mobility on different mobility scenarios if we examine the Interest transmission rates, for which the results are shown in Figure~\ref{fig:overhead_2AS}. With the Smart-flooding and Semi-flooding strategies, Interest transmission rate is barely affected by the \emph{Consumer} mobility, whereas for the \emph{mobile-producer} scenarios, we observe up to $\approx 10-30\%$ increase in overhead for the delay-sensitive traffic and up to $\approx 5-20\%$ for the delay-tolerant traffic. We observe higher overhead for the Flooding as well, but, the increase is mainly caused by the increased number of links compared to the single-AS scenario (note that, average hop-by-hop distance between end-points is still the same).



\section{Discussions}\label{Section:Discussions}

Simple flooding works reasonably well for both delay-sensitive and delay-tolerant traffic. However, the cost is significantly high, hence it does not scale well in network size and the number of hosts. On the other hand, we have Smart-flooding based strategies that are scalable but are not capable of providing the most basic servicing guarantees (keep in mind that, for interactive applications, one end point being mobile is sufficient to degrade the overall service quality).

It is possible to address mobility issues in Named Data Networking by exploiting the tradeoffs between these two scenarios. We can implement a specific mobility-based tagging mechanism for Data packets originated at mobile hosts to enable early Smart-flooding, for instance, by allowing the intermediate NDN routers to dynamically adapt the Interest timeout values. In doing so, content discovery process can finish sooner, since timeouts are used by the NDN routers to decide when to initiate a new search for Data packets targeted by the received Interests. However, at the same time, this may also increase the processing overhead at the NDN routers significantly because of forcing additional updates on PIT and FIB entries. Consequently, complexity increase at the NDN routers may not justify the performance improvements, as new forwarding paths will still need to be established, after each handover.

On the other hand, the above procedures cannot be implemented at the \emph{Consumer} side as resources at the NDN routers are expected to be shared by different \emph{Consumer}s that request the same content. Hence, to address the problems concerning \emph{Consumer} mobility, any solution starts at the \emph{Consumer} itself and, then, in its host network. For instance, \emph{Consumer} can implement an extended handover phase to take advantage of NDN's multi-homing feature, by simultaneously utilizing resources at the current access network and the next one. However, this process will continue to introduce additional overhead as some Data packets will still be delivered to the previous network and local resources at the NDN routers (\emph{i.e.}, active PIT entries, and Data packets at the Content Store) will continue to be allocated to such requests. We can reduce the resource usage requirements by allowing the \emph{Consumer} to optimally adjust the Interest message lifetime so as to enable the NDN routers free their local resources earlier, which, however, would require additional complexity at the host side.

In short, we observe that the current \emph{late-binding} architecture (\emph{i.e.}, resolving content requests to location using online forwarding strategies) utilized by NDN is not sufficient to handle mobility related problems. Furthermore, it is difficult to limit the messaging overhead associated with flooding after each handover. For these reasons, we need to investigate the use of \emph{iterative-binding} techniques (\emph{i.e.}, using name resolution servers that provide name-location mappings) in NDN. By integrating location-centric routing policies that separates name and network address into the current NDN architecture, we can enable quick recovery of named-data path without flooding the network for mobility-driven content delivery scenarios.

\section{Conclusions}\label{Section:Conclusions}
In this paper, we investigated the impact of \emph{Consumer} and \emph{Producer} mobility in information-centric networks by specifically focusing on the performance of NDN-based content delivery in wireless access networks. Through extensive simulations in ndnSIM using different mobility scenarios (\emph{i.e.}, intra-AS and inter-AS mobility) and application types (\emph{i.e.}, delay-sensitive and delay-tolerant traffic), we observed significant performance degradations in the effective throughput and the messaging overhead, pointing to the potential scalability problems caused by \emph{Consumer}/\emph{Producer} mobility in NDN. Even for small sized network topologies, NDN failed to satisfy the service quality requirements associated with each considered application type at different levels of host mobility without introducing significant overhead to the system. Our analysis showed the need to implement location-centric routing policies in NDN to satisfy the service quality needs of different user applications while minimizing the burden on the network infrastructure.

\bibliographystyle{IEEEtran}
\bibliography{ICN}

\end{document}